%% file: ProceedingTop2018.tex
\documentclass[12pt]{article}
\usepackage{graphicx}
\usepackage{url}

\newcommand\pubnumber{}
\newcommand\pubdate{\today}

\def\institute{on behalf of the ATLAS and CMS collaborations.}
\def\support{\footnote{University of Glasgow}}

\def\Title#1{\begin{center} {\Large #1 } \end{center}}
\def\Author#1{\begin{center}{ \sc #1} \end{center}}
\def\Address#1{\begin{center}{ \it #1} \end{center}}

\newcommand\pubblock{\rightline{\begin{tabular}{l} \pubnumber\\
         \pubdate  \end{tabular}}}
\newenvironment{Abstract}{\begin{quotation}  }{\end{quotation}}
\newenvironment{Presented}{\begin{quotation} \begin{center} 
             PRESENTED AT\end{center}\bigskip 
      \begin{center}\begin{large}}{\end{large}\end{center} \end{quotation}}

\input econfmacros.tex
\usepackage{lineno}
\usepackage{graphicx}
\usepackage{subcaption}
\begin{document}
%\linenumbers
\begin{titlepage}
\pubblock

\vfill
\Title{Comparative overview of differential measurements at ATLAS and CMS}
\vfill
\Author{ Federica Fabbri\support}
\Address{\institute}
\vfill
\begin{Abstract}
An overview of the measurements of differential top-quark pair production cross sections in proton-proton collisions at a center of mass energy of 13 TeV at the Large Hadron Collider using the ATLAS and CMS detectors is presented. Differential measurements obtained with respect to kinematic distributions of the top quark and   $t\bar{t}$ system, performed in different channels are discussed, with particular attention to the differences and similarities on the techniques employed and the results obtained from the two experiments.
These measurements probe our understanding of top quark production up to the TeV scale and, in general, show good agreement with the standard model expectations.
\end{Abstract}
\vfill
\begin{Presented}
$11^\mathrm{th}$ International Workshop on Top Quark Physics\\
Bad Neuenahr, Germany, September 16--21, 2018
\end{Presented}
\vfill
\end{titlepage}
\def\thefootnote{\fnsymbol{footnote}}
\setcounter{footnote}{0}

\section{Introduction}

The study of the top quark has a central role for both the ATLAS~\cite{ATLAS} and CMS~\cite{CMS} experiments. The measurement of the $t\bar{t}$ production differential cross section, in particular, is sensitive to existence of new resonances, it is a stringent test of perturbative QCD calculations and it is used to improve the modelling of $t\bar{t}$ production.
ATLAS and CMS provided a large number of results during the last years at $\sqrt{s}$=7, 8, 13 TeV, in different regions of the phase space, considering different channels and topology and measuring the cross section as a function of the kinematic variables of the $t\bar{t}$ system, the top quarks and/or their decay products.\\
Even if the individual steps are dependent on the specific analysis all the differential cross section measurements proceed through the same workflow: the event selection, the background determination, the reconstruction of the $t\bar{t}$ system, the definition of the fiducial phase space, the unfolding applied to remove the effect of limited acceptance and resolution of the detector and the evaluation of the systematic uncertainties.
In the following I will focus on the techniques used by ATLAS and CMS to perform some of these steps while presenting a selection of recent results by the two experiments.

\section{Dilepton channel}
\begin{figure}[htb]
\centering
\begin{subfigure}[t]{0.42\textwidth}
\centering
\includegraphics[width=0.93\textwidth]{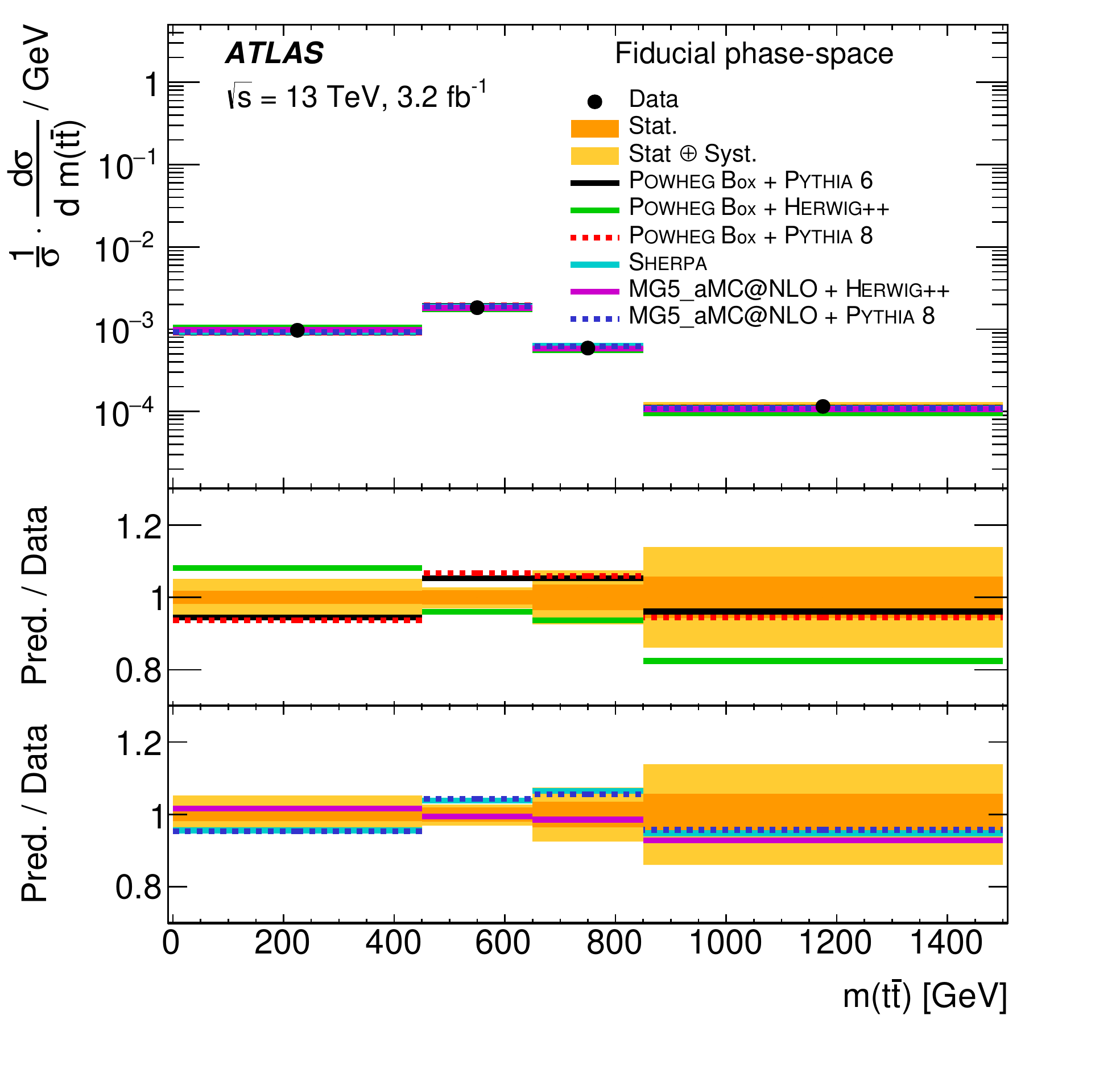}
\end{subfigure}
\begin{subfigure}[t]{0.42\textwidth}
\centering
\includegraphics[width=\textwidth]{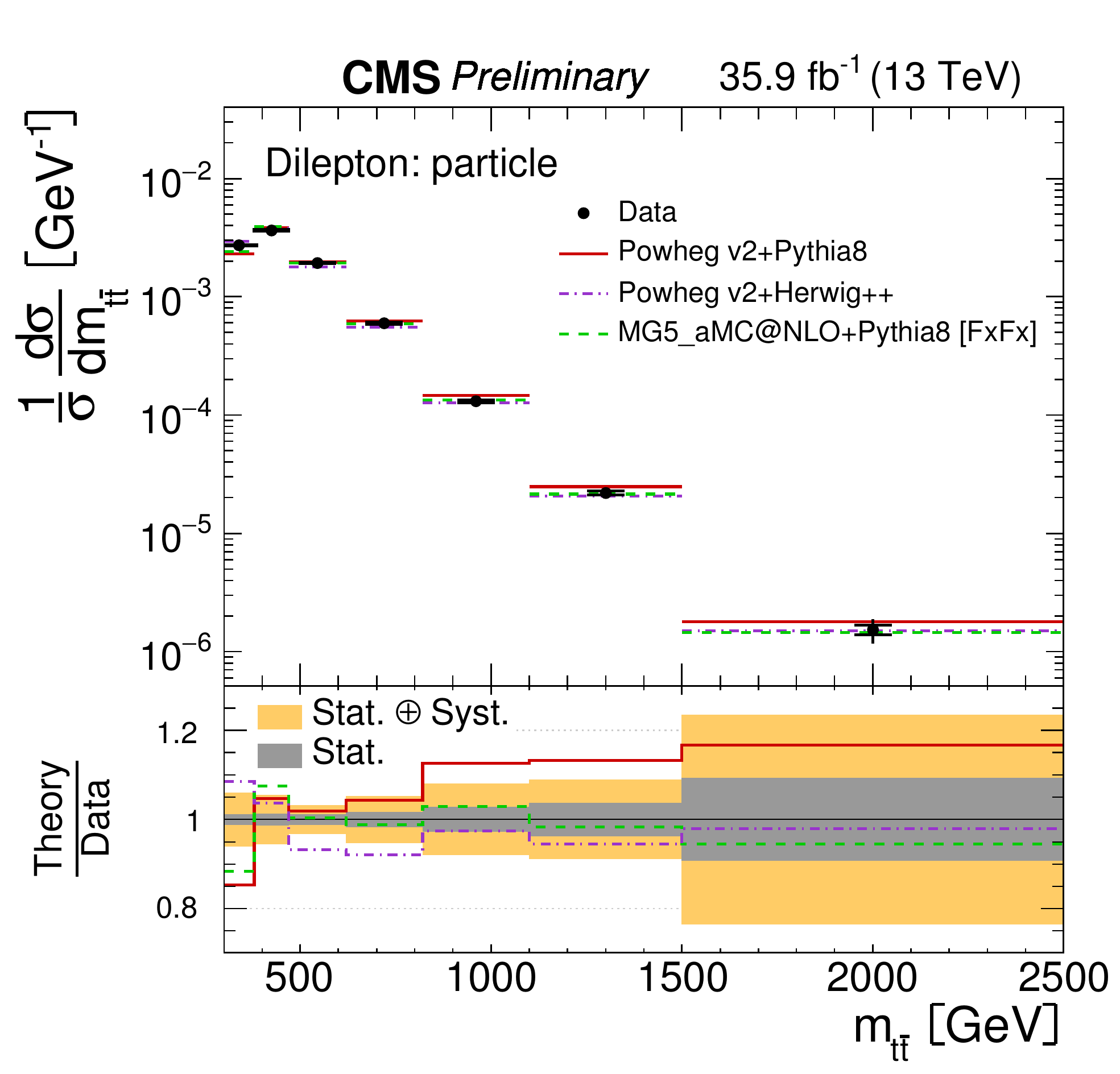}
\end{subfigure}
\caption{$\frac{1}{\sigma}\frac{d\sigma_{t\bar{t}}}{d m_{t\bar{t}}}$ at particle level measured by ATLAS~\cite{ATLAS-dil} (left) and CMS~\cite{CMS-dil} (right), at $\sqrt{s}=$13 TeV compared with NLO predictions. The yellow band corresponds to the sum of systematic and statistical uncertainties.}
\label{fig:dilep}
\end{figure}

The dilepton channel is characterized by the presence of two leptons, two $b$-jets and large $E_{\mathrm{T}}^{\mathrm{miss}}$ in the final state. The main challenge of the differential cross section measurement in this channel is to reconstruct the six components of the two neutrino ($\nu$) momenta in the events, starting only from the measured $E_{\mathrm{T}}^{\mathrm{miss}}$. In both ATLAS and CMS approaches, the mass of the $W$ boson and top quark are added as constraints to perform the $t\bar{t}$ reconstruction.
In the CMS approach, for each event 100 trials are obtained smearing the reconstructed quantities of leptons and $b$-jets within their resolution and the $\nu$ kinematics are evaluated for each trial imposing the four-momentum conservation. The final $\nu$ momenta are then obtained as weighted average among all the trials where a solution was found, considering the truth mass of the lepton - $b$-jet system in the weights. This technique allows to reconstruct the $t\bar{t}$ system with a 90$\%$ efficiency on the selected signal. In the ATLAS approach a scan is performed using several values of $\eta^{\nu}$ and $\eta^{\bar{\nu}}$ to solve for the $\nu$ momenta and a weight containing the difference between the estimated and measured $E_{\mathrm{T}}^{\mathrm{miss}}$ is assigned to each solution. The final $\nu$ is the solution with the highest weight. This technique has a reconstruction efficiency of $\approx$ 80$\%$ on the selected signal. The approach used to define the particle level $t\bar{t}$ is the same for the two measurements and is based on finding the combination that minimizes $|W_{1,reco}^{mass} -W^{mass}|+|W_{2,reco}^{mass} -W^{mass}|$, and the same equation considering the top quark mass.
Figure~\ref{fig:dilep} shows the $\frac{1}{\sigma}\frac{d\sigma_{t\bar{t}}}{d m_{t\bar{t}}}$ measured by ATLAS~\cite{ATLAS-dil} and CMS~\cite{CMS-dil} at $\sqrt{s}$=13, TeV compared with several next to leading (NLO) predictions. Both experiments obtain a fair agreement with the predictions and observe the same trend in the agreement with Powheg\cite{Frixione:2007nw} interfaced with Herwig++\cite{Herwig++}.
In both measurements the leading uncertainties are due to the modelling of $t\bar{t}$ production.

\section{Lepton+jets channel}

\begin{figure}[!h]
\centering
\begin{subfigure}[t]{0.42\textwidth}
\centering
\includegraphics[width=0.95\textwidth]{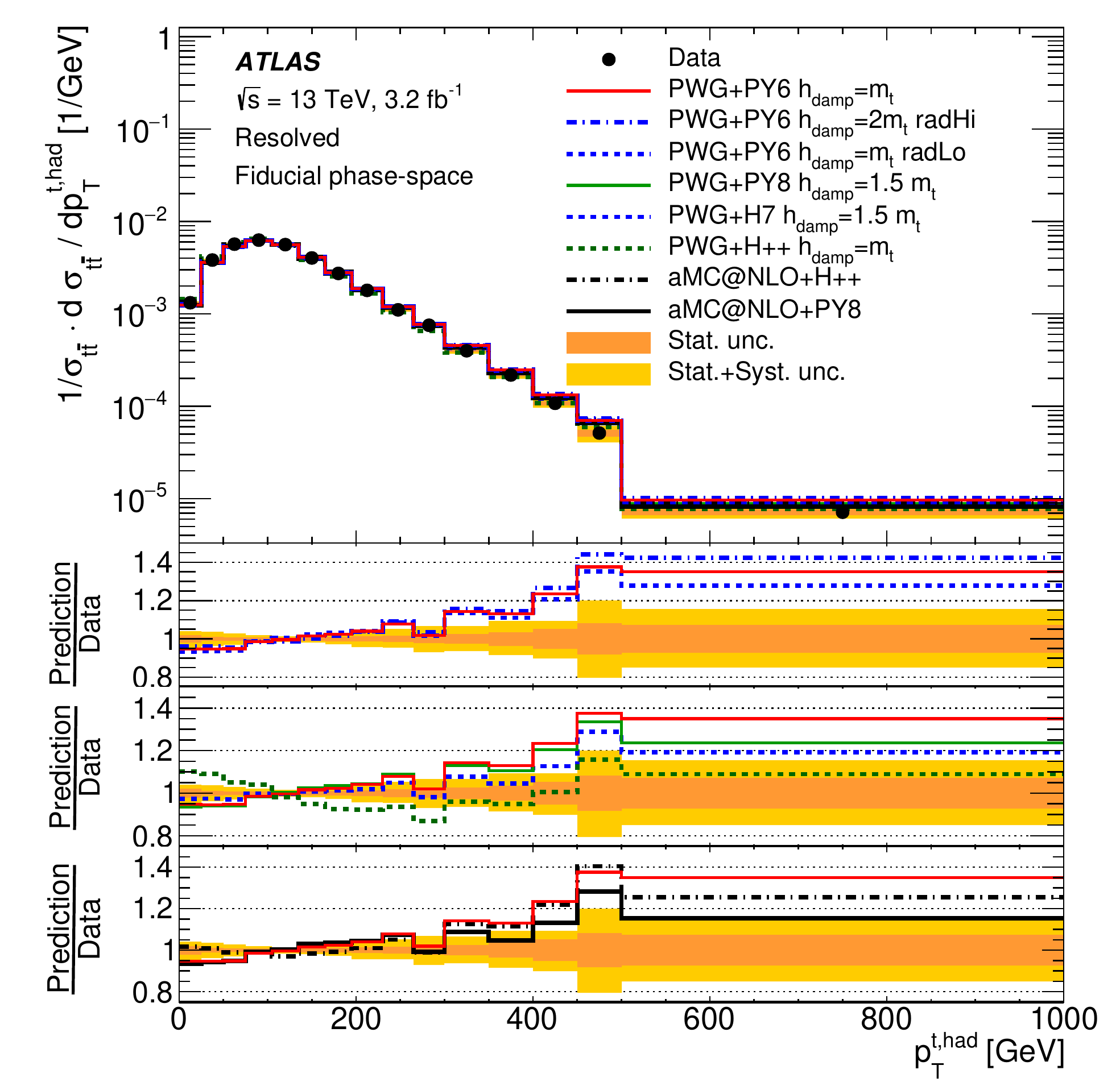}
\end{subfigure}
\begin{subfigure}[t]{0.42\textwidth}
\centering
\includegraphics[width=\textwidth]{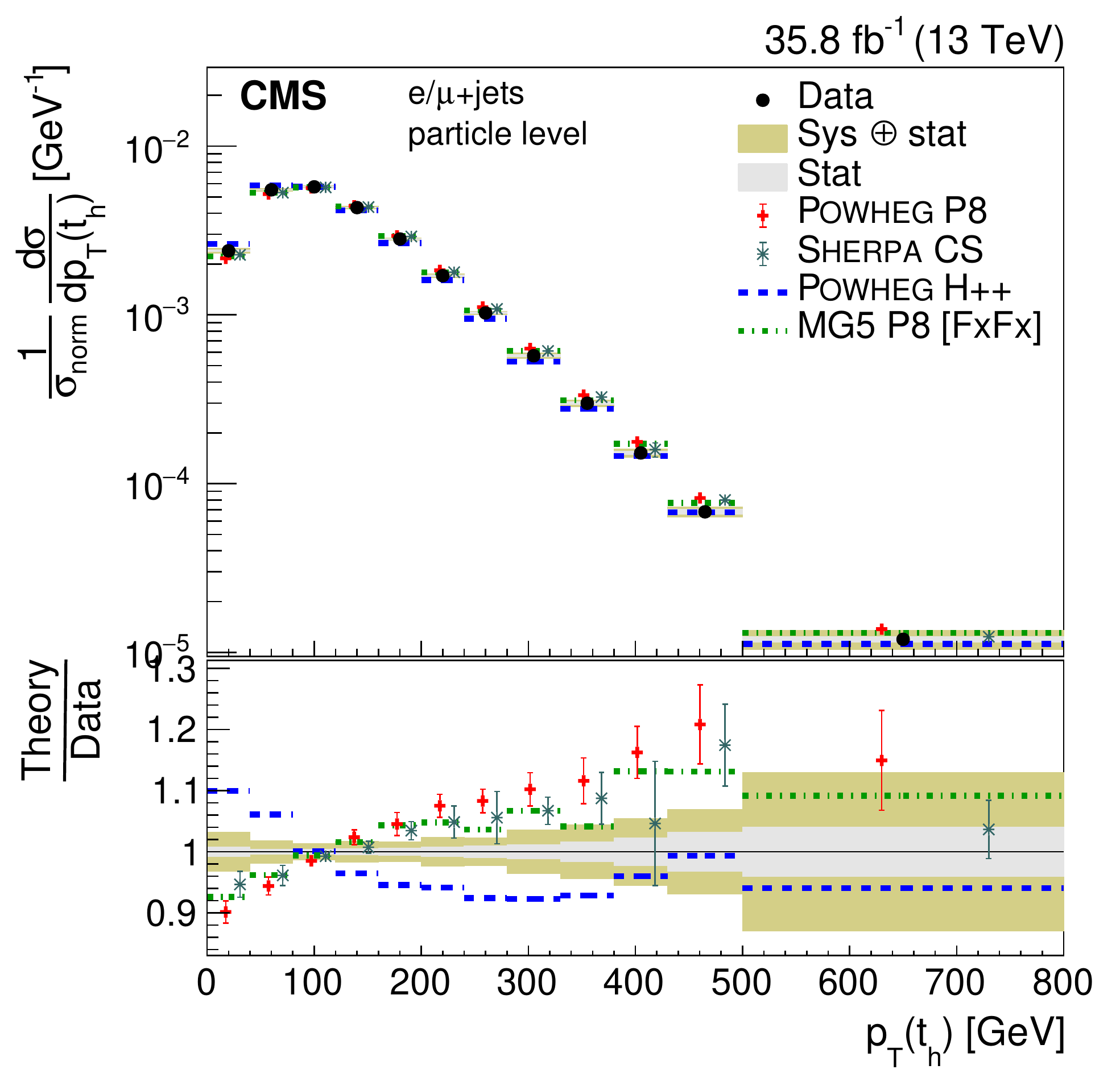}
\end{subfigure}

\begin{subfigure}[t]{0.42\textwidth}
\centering
\includegraphics[width=0.95\textwidth]{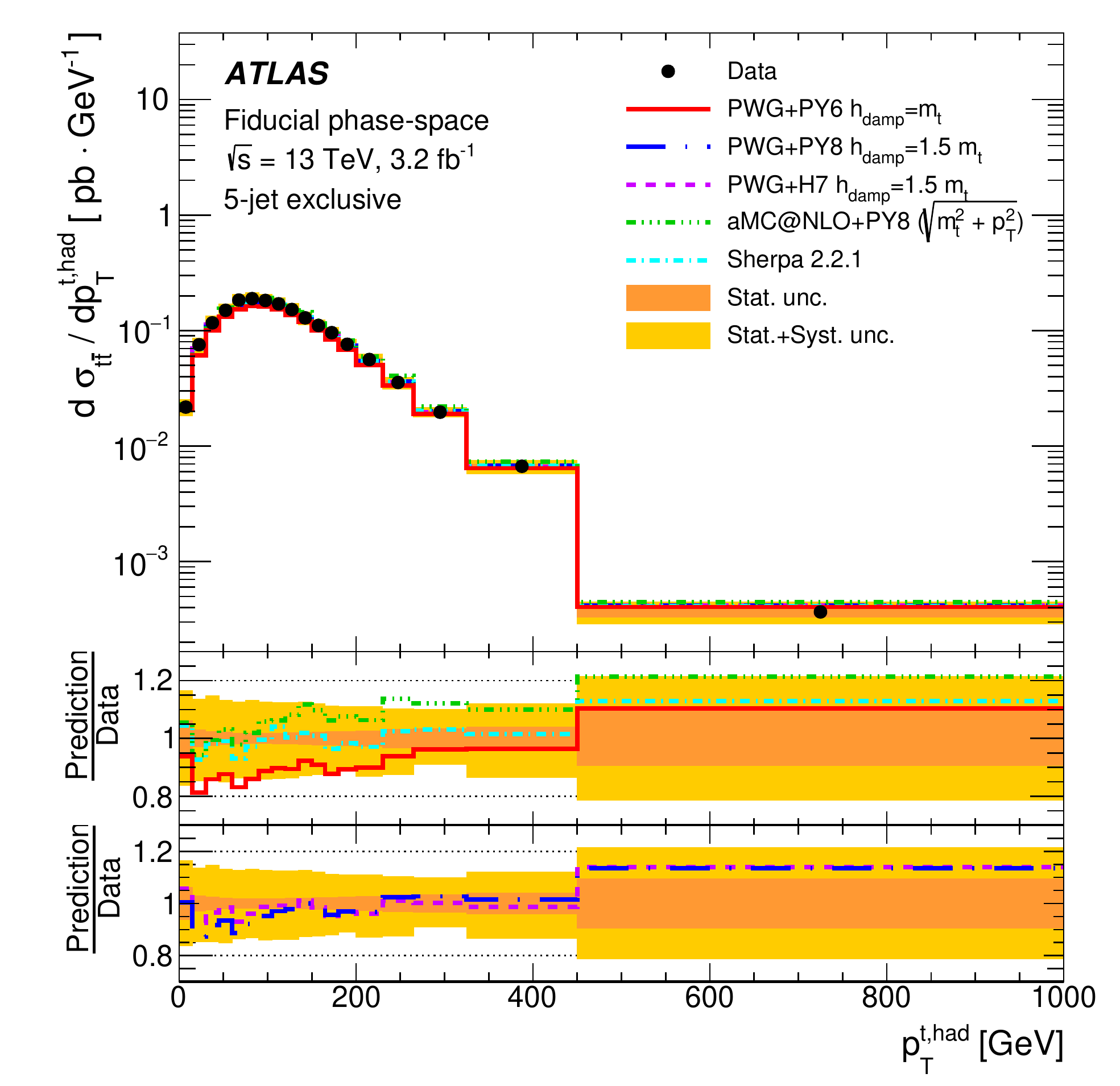}
\end{subfigure}
\begin{subfigure}[t]{0.42\textwidth}
\centering
\includegraphics[width=\textwidth]{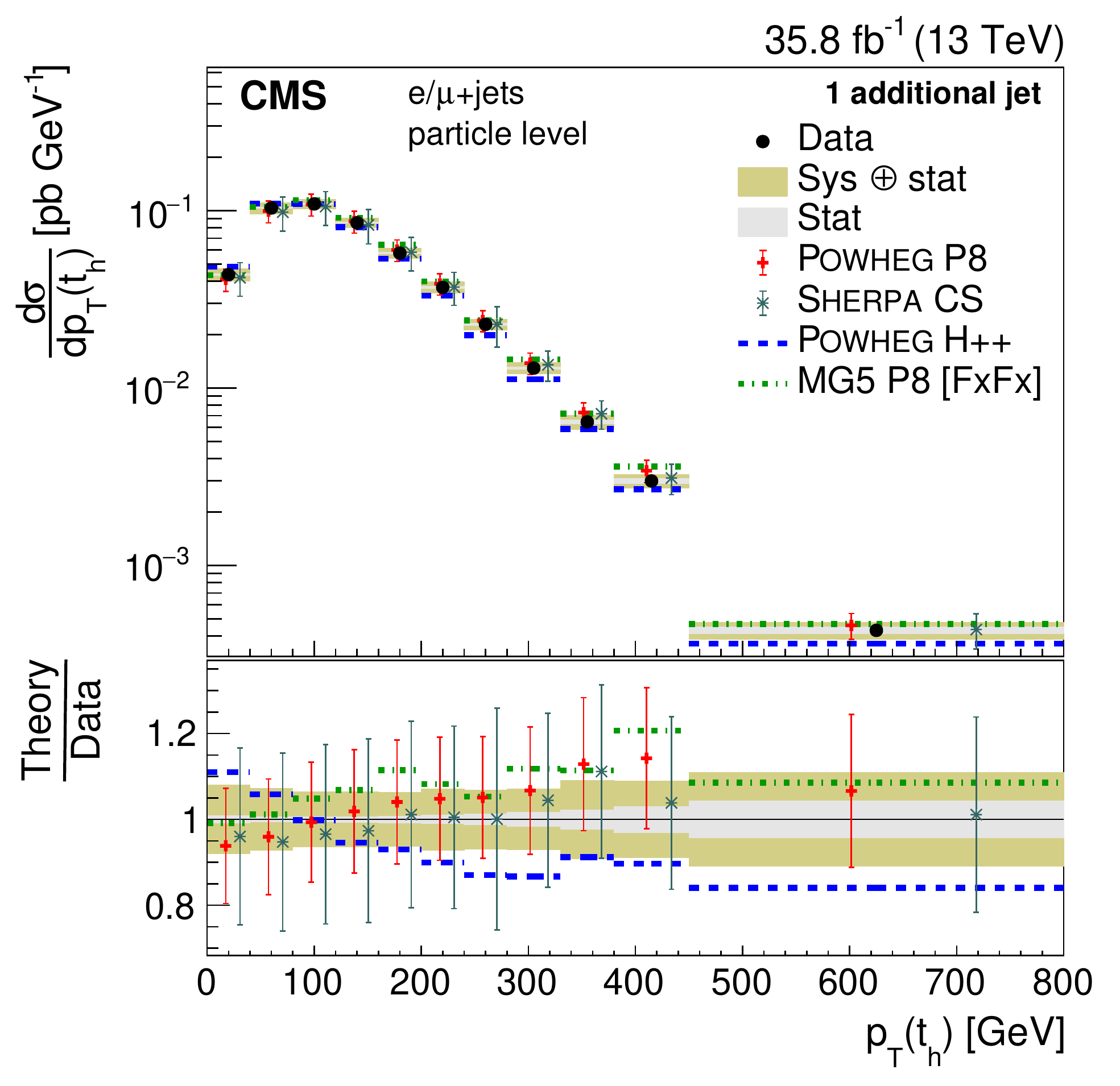}
\end{subfigure}
\caption{$\frac{1}{\sigma}\frac{d\sigma_{t\bar{t}}}{d m_{t\bar{t}}}$ at particle level measured by ATLAS~\cite{ATLAS-ljet}~\cite{ATLAS-ljet-5j} (left) and CMS~\cite{CMS-ljet} (right), at $\sqrt{s}=$13 TeV compared with NLO predictions. The top row shows the cross section inclusive in number of additional jets, while in the second row are considered only events with exactly 1 additional jet.}
\label{fig:ljets}
\end{figure}

The lepton+jets channel final state includes at least 4 jets, two of which originating from $b$-quarks, one lepton and one neutrino. The correct reconstruction of the $t\bar{t}$ system needs to address the potential combinatorial errors. Both ATLAS and CMS measurements employ the pseudo-top ~\cite{ATLAS-pseudotop}\cite{CMS-pseudotop} algorithm at particle level to reconstruct the $t\bar{t}$ system. At detector level the ATLAS measurements apply exactly the same algorithm, with the advantage of obtaining small migrations in the response matrices connecting detector and particle level. The approach followed by CMS at detector level, instead, is based on the construction of probability functions obtained using MC simulations~\cite{CMS-ljet}.
The dominant uncertainties for both the measurements are due to the jet energy scale, in the low $p_{\mathrm{T}}^{t}$ ($<$ 400 GeV) region, and to the $t\bar{t}$ production modelling uncertainty in the high $p_{\mathrm{T}}$ one. In the ATLAS measurements the flavour tagging uncertainty also plays a significant role.\\
The $\frac{1}{\sigma}\frac{d \sigma_{t\bar{t}}}{d p_{\mathrm{T}}^{t}}$ obtained by ATLAS~\cite{ATLAS-ljet} and CMS~\cite{CMS-ljet} at particle level, compared with several NLO predictions, are shown in Figure~\ref{fig:ljets} (top row). The two measurements observe a trend in the agreement with the data: the NLO predictions seems to overestimate the data in the high $p_{\mathrm{T}}$ tails. The agreement between predictions and data is consistent between the two measurements.
An even more differential measurement of this distribution can be obtained evaluating the differential cross section in bins of additional jets, defined as jets not used to reconstruct the ${t\bar{t}}$ system. For example, in Figure~\ref{fig:ljets} (bottom row) is shown the $\frac{1}{\sigma}\frac{d\sigma_{t\bar{t}}}{d p_{\mathrm{T}}^{t}}$ obtained by ATLAS~\cite{ATLAS-ljet-5j} and CMS~\cite{CMS-ljet} with exactly one additional jet. In this case both measurements show an improvement in the agreement with the predictions with respect to the inclusive case. 

\section{Boosted topology}

\begin{figure}[hb!]
\centering
\begin{minipage}{0.42\textwidth}
\centering
\includegraphics[width=0.93\textwidth]{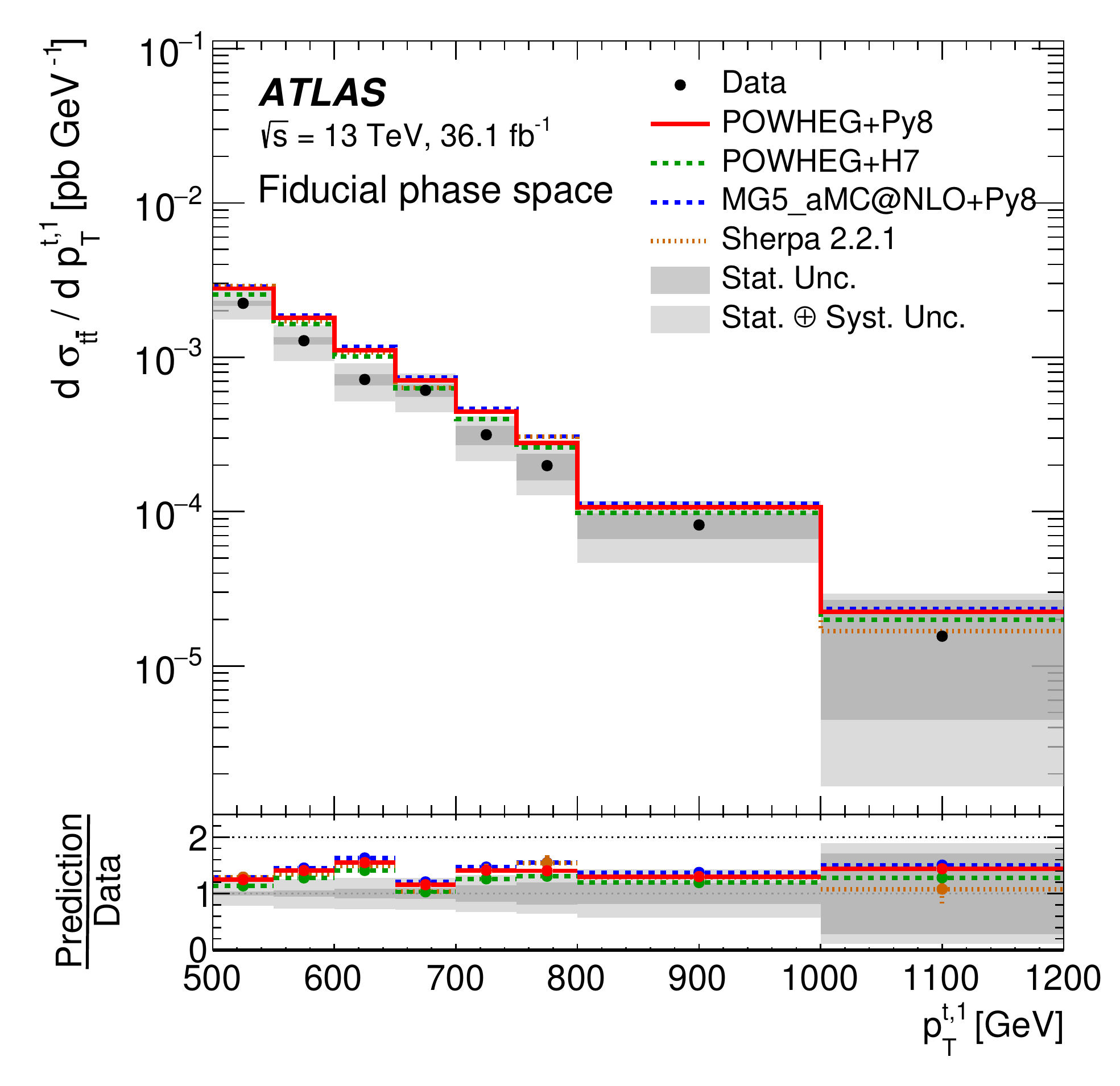}
\end{minipage}
\begin{minipage}{0.45\textwidth}
\centering
\includegraphics[width=\textwidth]{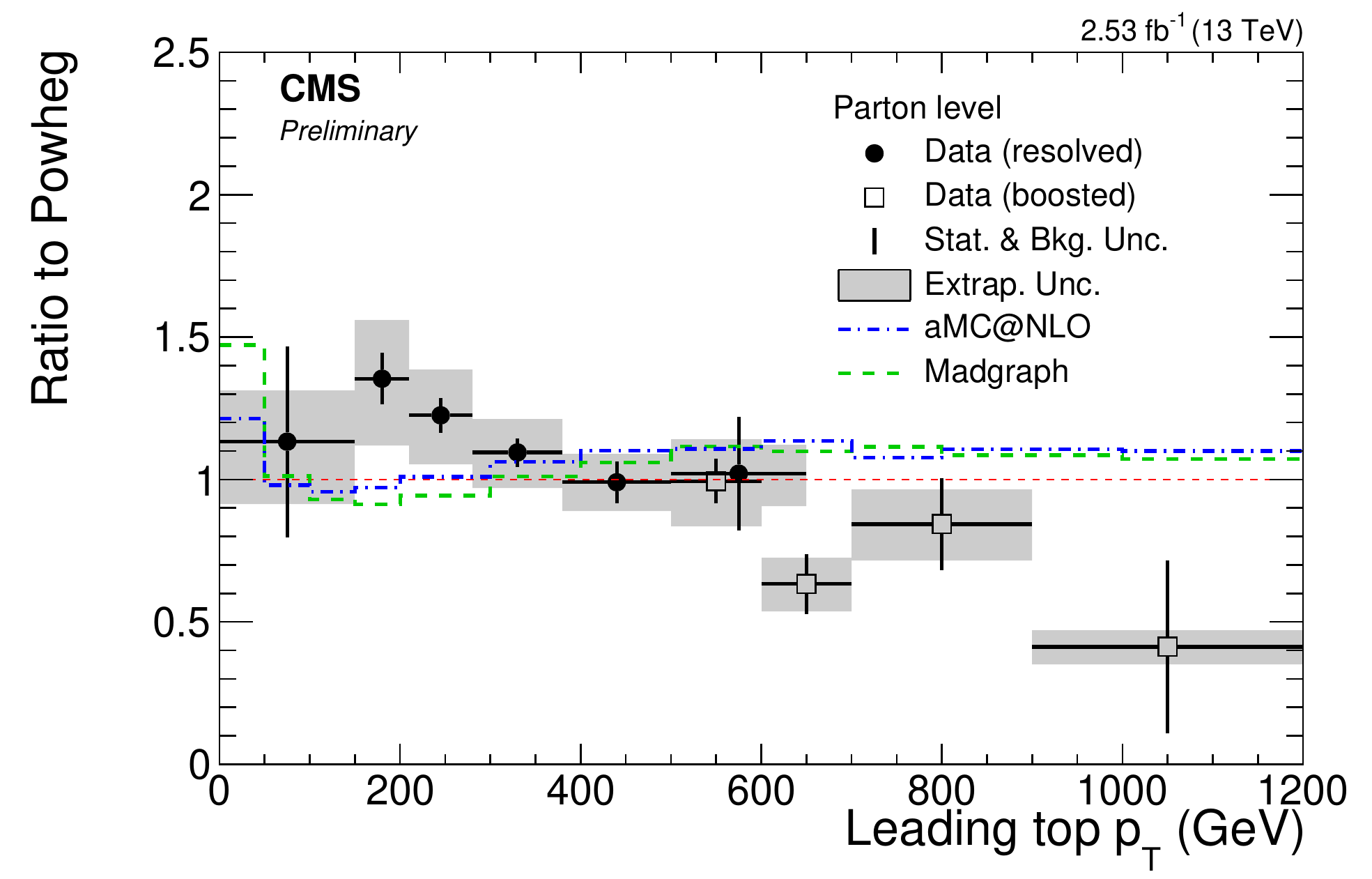}
\end{minipage}
\caption{$\frac{d \sigma_{t\bar{t}}}{d p_{\mathrm{T}}^{t, leading}}$ measured by ATLAS~\cite{ATLAS-alljet} at particle level (left) and CMS~\cite{CMS-alljet} at parton level (right), at $\sqrt{s}=$13 TeV. In the left plot the measured cross section is compared with many NLO predictions while on the right is presented the ratio between the measurement and Powheg through the full  $p_{\mathrm{T}}^{t, leading}$ spectra.}
\label{fig:boo}
\end{figure}

The high energy provided by the Large Hadron Collider allows to explore the frontier kinematic regime at the TeV scale. If the top quark is produced at very high $p_{\mathrm{T}}$, the decay products get a large lorentz boost and are all contained in a cone with radius $\approx \frac{2m}{p_{\mathrm{T}}}$.
In these cases only one large-radius jet is used to reconstruct the hadronically decaying top and it is identified thanks to substructure variables.
Both ATLAS~\cite{ATLAS-alljet} and CMS~\cite{CMS-alljet} performed measurements in this high $p_{\mathrm{T}}$ regime in the all-hadronic channel, the results on $\frac{d\sigma_{t\bar{t}}}{d p_{\mathrm{T}}^{t,leading}}$ are shown in Figure~\ref{fig:boo}. Both measurements see a general overestimate of the data in this phase space, consistent with observations in other channels.

\section{Conclusions}
\begin{figure}[ht!]
\centering
\includegraphics[width=0.35\textwidth]{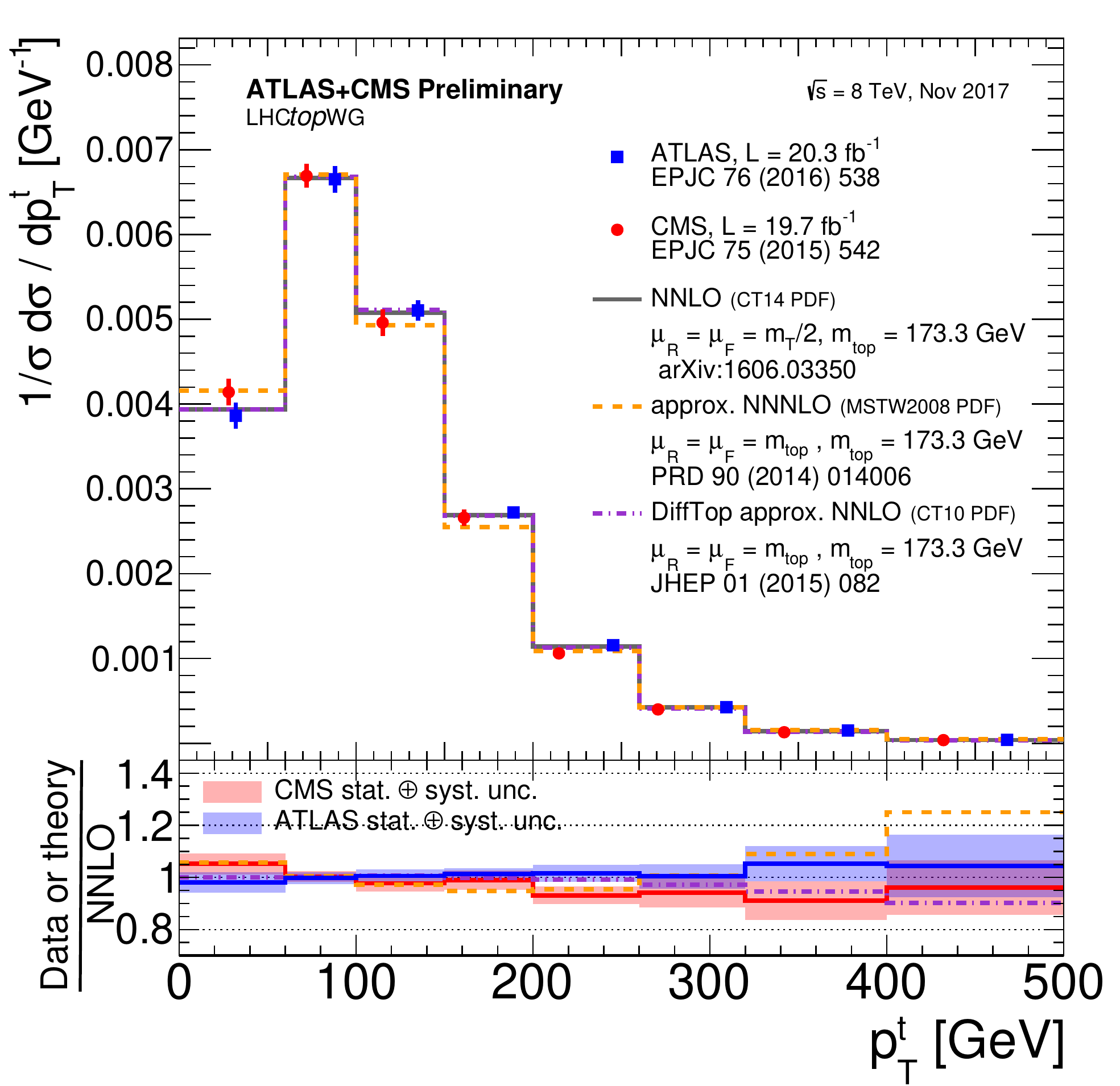}
\caption{ATLAS and CMS parton level measurements at $\sqrt{s}$= 8 TeV compared with NNLO calculation.}
\label{fig:tlhcwg}
\end{figure}

\footnote{Copyright 2019 CERN for benefit of the ATLAS and CMS collaboration. CC-BY-4.0 licence.} ATLAS and CMS presented a significant number of differential measurements of ${t\bar{t}}$ production. In general, the particle level results show good agreement with the NLO predictions, although the predictions are seem to tend to overestimate the data at high $p_{\mathrm{T}}^{t}$ values in all channels.
The important next step for differential measurements in the two experiments is to analyse the full dataset now collected by LHC, corresponding to ~140 $\mathrm{fb}^{-1}$ and provide quantitative comparison among the results obtained, repeating what was done at $\sqrt{s}$= 8 TeV, shown in Figure~\ref{fig:tlhcwg}. Here the two parton level results of $\frac{d \sigma_{t\bar{t}}}{d p_{\mathrm{T}}^{t}}$ performed by ATLAS and CMS were compared to the same NNLO theoretical calculation. Moreover, it is necessary to extend the comparison to particle level and obtain an LHC combination.
This effort is very relevant to maximize the information that can be extracted from the differential measurements in terms of testing the QCD calculations, search for new physics and improve the description of the parton distribution functions of the gluons inside the protons.

%%%%%%%%%%%%%%%%%%%%%%%%%%%%%%%%%%%%%%%%%%%%%%%%%%%%%%%%%%%%%%%%%%%%%%%%%%
%%%
%%%   use this format to include a LaTeX table  into your paper
%%%
%\begin{table}[t]
%\begin{center}
%\begin{tabular}{l|ccc}  
%Patient &  Initial level($\mu$g/cc) &  w. Magnet &  
%w. Magnet and Sound \\ \hline
% Guglielmo B.  &   0.12     &     0.10      &     0.001  \\
% Ferrando di N. &  0.15     &     0.11      &  $< 0.0005$ \\ \hline
%\end{tabular}
%\caption{Blood cyanide levels for the two patients.}
%\label{tab:blood}
%\end{center}
%\end{table}
%%%%%%%%%%%%%%%%%%%%%%%%%%%%%%%%%%%%%%%%%%%%%%%%%%%%%%%%%%%%%%%%%%%%%%%%%%%
%

\end{document}

%% file: econfmacros.tex
\def\beq{\begin{equation}}
\def\eeq#1{\label{#1}\end{equation}}
\def\eeqn{\end{equation}}

\def\beqa{\begin{eqnarray}}
\def\eeqa#1{\label{#1}\end{eqnarray}}
\def\eeqan{\end{eqnarray}}

\let\bar=\overbar

\def\Dslash{\not{\hbox{\kern-4pt $D$}}}
\def\dslash{\not{\hbox{\kern-2pt $\del$}}}

\def\msb{{\bar{\ssstyle M \kern -1pt S}}}